\documentclass[11pt,twoside]{article}

\usepackage{asp2006}
\usepackage{epsf}
\usepackage{psfig}
\usepackage{lscape}
\usepackage{graphicx}

\newcommand{\mj}{$M_{\mathrm{J}}$}
\newcommand{\rj}{$R_{\mathrm{J}}$}
\newcommand{\me}{$M_{\oplus}$}
\newcommand{\re}{$R_{\oplus}$}

\newcommand{\hd}{HD 209458b} 
\newcommand{\hh}{HD 149026b}

\newcommand{\cp}{\citep}
\newcommand{\ct}{\citet}

\begin{document}
\title{The Impact of Transit Observations on Planetary Physics}   
\author{Jonathan J. Fortney}   
\affil{NASA Ames Research Center, and Department of Astronomy \& Astrophysics, University of California, Santa Cruz}    

\begin{abstract} 
We highlight the importance of transit observations on understanding the physics of planetary atmospheres and interiors.  Transmission spectra and emission spectra of atmospheres allow us to characterize these exotic atmospheres, which possess TiO, VO, H$_2$O, CO, Na, and K, as principal absorbers.  We calculate mass-radius relations for water-rock-iron and gas giant planets and examine these relations in light of current and future transit observations.  A brief review is given of mechanisms that could lead to the large radii observed for some transiting planets.
\end{abstract}


\section{Introduction}   
\begin{figure}[!ht]
\begin{center}
\includegraphics[scale=0.62]{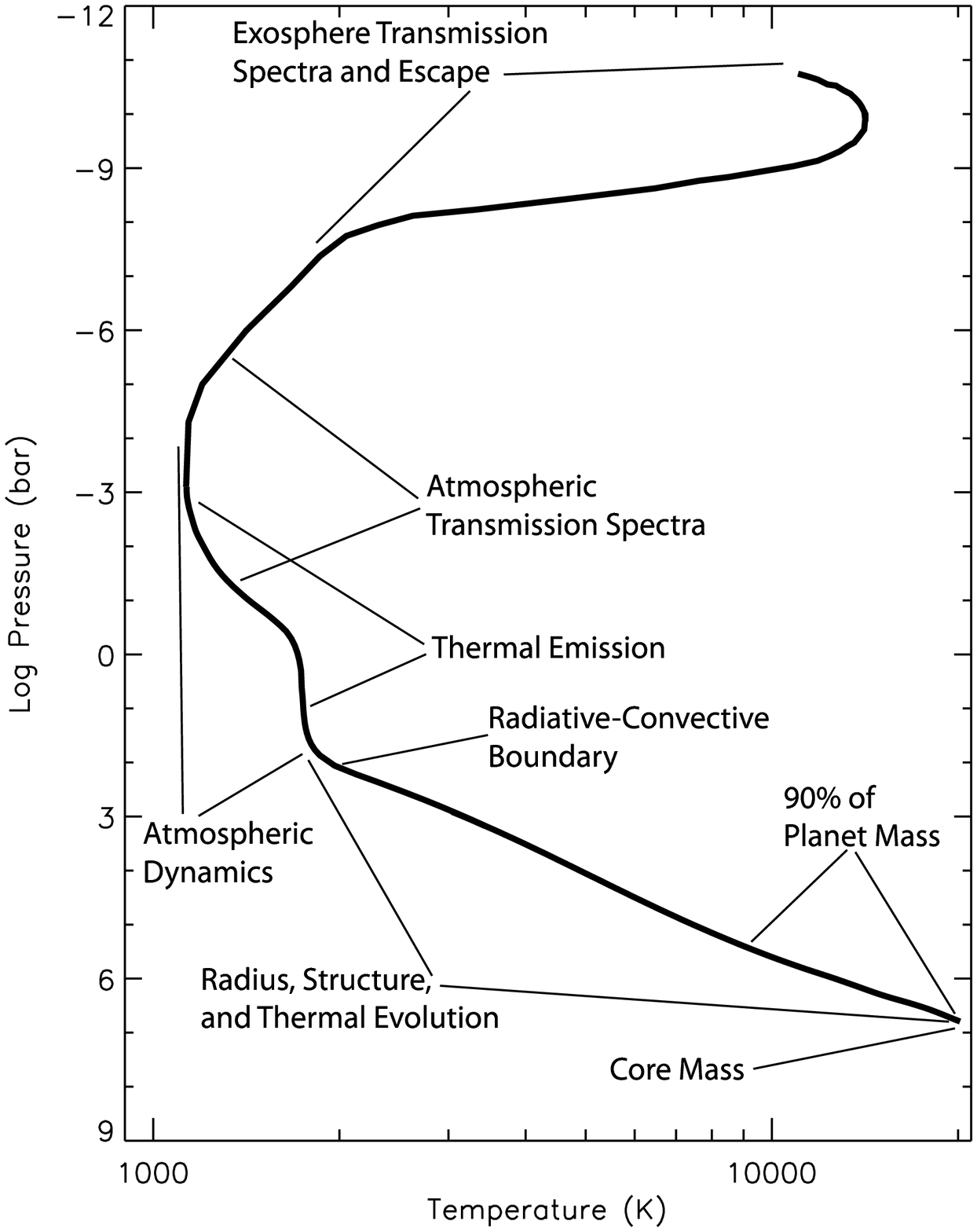}
\end{center}
\caption{Exosphere \cp[adapted from][]{Yelle04} , atmosphere \citep{Fortney05}, and interior \cp{Fortney07a} \emph{P-T} profile for a generic HD 209458b-like planet.}\label{moth}
\end{figure}

The blanket term ``hot Jupiter" or even the additional term ``very hot Jupiter'' belies the diversity of these highly irradiated planets.  Each planet likely has its own unique atmosphere, interior structure, and accretion history.  The relative amounts of refractory and volatile compounds in a planet will reflect the parent star abundances, nebula temperature, total disk mass, location of the planet's formation within the disk, duration of its formation, and its subsequent migration (if any).  This accretion history will give rise to differences in core masses, total heavy elements abundances, and atmospheric abundance ratios.  Through transit observations we are able to probe nearly \emph{20 orders of magnitude} in pressure in these planets, as seen in \mbox{Figure~\ref{moth}}, from the outer reaches of the escaping atmosphere to the high pressures of dense central cores.

Since atmospheric and interior physics are both quite large topics, here we will focus specifically on the atmospheres of close-in giant planets at photospheric pressures, which \emph{Spitzer} data is allowing us access to, as well as the structure and mass-radius relations of Earth-like to gas giant planets.  At least 21 gas giants are now known, as well as a possible Neptune-class planet, with even smaller planets on the horizon.
\begin{figure}[!ht]
\begin{center}
\includegraphics[scale=0.53]{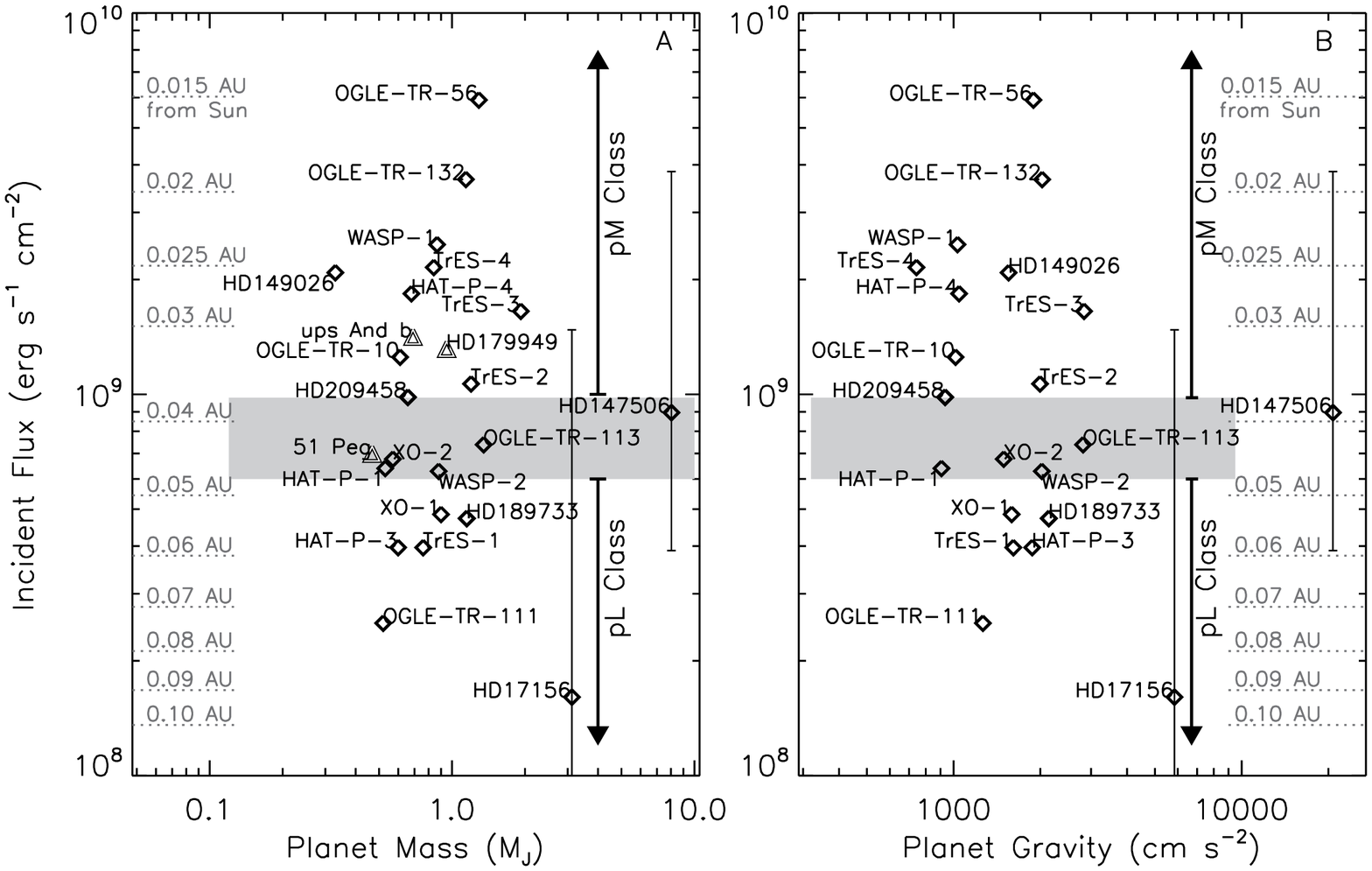}
\end{center}
\caption{Flux incident upon a collection of hot Jupiter planets.  At left is incident flux as a function of planet mass, and at right as a function of planet surface gravity.  In both figures the labeled dotted lines indicate the distance from the Sun that a planet would have to be to intercept this same flux.  Diamonds indicate the transiting planets while triangles indicate non-transiting systems  (with minimum masses plotted but unknown surface gravities).  The error bars for HD 147506 and HD17156 indicate the variation in flux that each planet receives over their eccentric orbits.  Flux levels for pM Class and pL Class planets are shown, with the shaded region around $\sim$0.04-0.05 AU indicating a possible transition region.  ``Hot Neptune'' GJ 436b experiences less intense insolation, off the bottom of this plot at 3.2$\times$10$^7$ erg s$^{-1}$ cm$^{-2}$.}\label{flux}
\end{figure}
\section{Atmospheres}
Since irradiation is perhaps the most important factor in determining the atmospheric properties of these planets, we examine the insolation levels of the 20 known transiting planets.  We restrict ourselves to those planets more massive than Saturn, and hence for now exclude treatment of the ``hot Neptune'' GJ 436b.  \mbox{Figure~\ref{flux}} illustrates the stellar flux incident upon the planets as a function of both planet mass (\mbox{Figure~\ref{flux}}\emph{a}) and planet surface gravity (\mbox{Figure~\ref{flux}}\emph{b}).  In these plots diamonds indicate transiting planets and triangles indicate other interesting hot Jupiters, for which \emph{Spitzer Space Telescope} data exist, but which do not transit.  The first known transiting planet, \hd, is seen to be fairly representative in terms of incident flux.

\ct{Fortney07c} argue that based on the examination of few physical processes that two classes of hot Jupiter atmospheres emerge.  Equilibrium chemistry, the depth to which incident flux will penetrate into a planet's atmosphere, and the radiative time constant as a function of pressure and temperature in the atmosphere all naturally define two classes these irradiated planets.   Those planets that are warmer than required for condensation of titanium (Ti) and vanadium (V)-bearing compounds will possess a temperature inversion at low pressure due to absorption of incident flux by TiO and VO (a stratosphere, as found by \citealt{Hubeny03} and \citealt{Fortney06}), and will appear ``anomalously'' bright in secondary eclipse at mid-infrared wavelengths.  Furthermore, these planets will have large day/night effective temperature contrasts \emph{and} will also be comparatively easy to detect at optical wavelengths due to thermal emission.  We will term these very hot Jupiters the ``pM Class,'' meaning gaseous TiO and VO are the prominent absorbers of optical flux, similar to M dwarfs.  Planets with temperatures below the condensation temperature curve of Ti and V bearing compounds will have relatively smaller secondary eclipse depths in the mid infrared, significantly smaller day/night effective temperature contrasts, and atmospheric dynamics will lead to a complex redistribution of absorbed energy.  We term these hot Jupiter planets the ``pL Class,'' since once TiO/VO are removed, alkalis become the prominent optical absorbers \cp*{BMS}, similar to L dwarfs.  Published \emph{Spitzer} data are consistent with this picture.  In the late stages of the submission of this contribution, \ct{Knutson08} used \emph{Spitzer} IRAC to find evidence for a temperature inversion for \hd\ at secondary eclipse, potentially attributed to TiO/VO opacity \cp{Burrows07c}.  The inversions for planets \hd\ \cp{Knutson08} and \hh\ \cp[][as was predicted by \citealp{Fortney06}]{Harrington07} support the theory of \ct{Fortney07c}.  \citet{Burrows08} have recently advanced a similar view.

\begin{figure}[!ht]
\begin{center}
\includegraphics[scale=0.70]{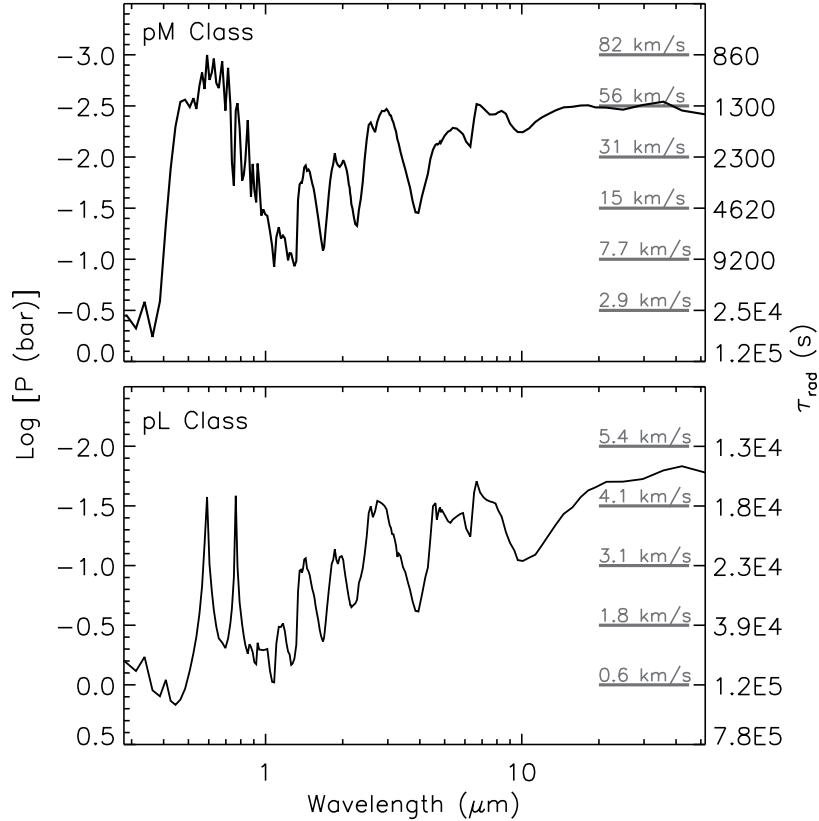}
\end{center}
\caption{Both figures show, as a function of wavelength, the pressure that corresponds to a calculated brightness temperature.  Note the differences in the y-axes.  \emph{Top}: Planet at 0.03 AU which has a stratosphere induced by absorption by TiO/VO.  \emph{Bottom}: Planet at 0.05 AU that lacks a stratosphere.  The right ordinate (which i not linear) shows the corresponding radiative time constant at each major tick mark from the pressure axis.  The labeled gray lines at right indicate an advective wind speed that would be necessary to give an advective time scale equal to the given radiative constant.}\label{ptau}
\end{figure}

A key difference between the atmospheres of the pL Class planets and pM Class planets is the pressure at which the absorption and emission of flux occurs \cp{Fortney07c}.  This can be shown by plotting the pressure that corresponds to a calculated brightness temperature.  This is shown as a function of wavelength in \mbox{Figure~\ref{ptau}}.  In general thermal emission arises from a characteristic pressure level roughly an order of magnitude greater in a pL Class atmosphere than in a pM Class atmosphere.

\begin{figure}[!ht]
\begin{center}
\includegraphics[scale=0.65]{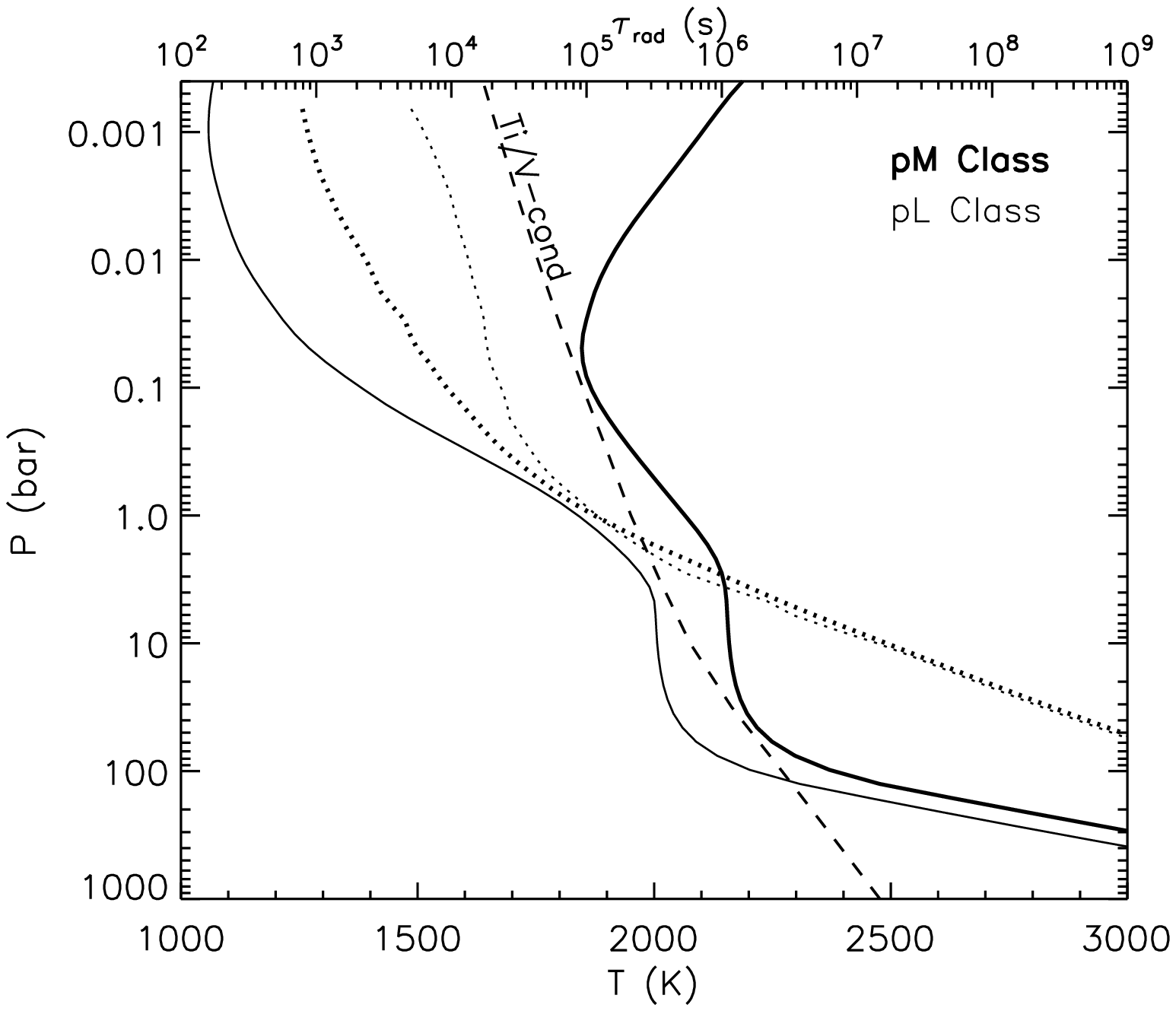}
\end{center}
\caption{Pressure-temperature (solid lines, bottom x-axis) and pressure-$\tau_{rad}$ profiles (dotted lines, top x-axis) for the models at 0.03 AU (pM Class, thick lines) and 0.05 AU (pL Class, thin lines).  Dashed curve shows Ti/V condensation.}\label{PTtau}
\end{figure}

What effect this may have on the dynamical redistribution of energy in a planetary atmosphere can be considered after calculation of the radiative time constant, $\tau_{\rm rad}$.  In the Newtonian cooling approximation a temperature disturbance relaxes exponentially toward radiative equilibrium with a characteristic time constant.  At photospheric pressures this value can be approximated by
\begin{equation}
\label{trad1}
\tau_{\rm rad} \sim \frac{P}{g} \frac{c_{\rm P}}{4 \sigma T^3},
\end{equation}
where $\sigma$ is the Stefan-Boltzmann constant and $c_{\rm P}$ is the specific heat capacity \cp{Showman02}.  
In \mbox{Figure~\ref{PTtau}} we show detailed calculations of $\tau_{\rm rad}$ as a function of pressure alongside the radiative-convective equilibrium profiles from which they were generated \cp{Fortney07c}.  Although these time constants are nearly equivalent in the dense lower atmosphere, they are significantly different in the thinner upper atmosphere that one is sensitive to from mid-infrared observations.

The right y-axis in \mbox{Figure~\ref{ptau}} is cast in terms of the radiative time constant appropriate for a given pressure in the atmosphere.  While those for the pM Class range from only $10^3-10^4$ s, those for the pL Class range from $10^4-10^5$ seconds.  This $\sim 10 \times$ difference in timescale is a consequence of both the hotter temperatures and lower pressures of the pM Class photospheres, as can be understood from \mbox{Equation~\ref{trad1}}.

One can also define an advective time scale, a characteristic time for moving atmospheric gas a given planetary distance.  A common definition is \mbox{$\tau_{\rm advec} = \frac{R_{\rm p}}{U}$}, where $U$ is the wind speed and $R_{\rm p}$ is the planet radius \cp{Showman02,Seager05}.  If one sets $\tau_{\rm advec}=\tau_{\rm rad}$ and sets $R_{\rm p}=1 R_{\rm J}$, one can derive the wind speed $U$ that would be necessary to advect atmospheric gas before a $\tau_{\rm rad}$ has elapsed.  This is also shown in \mbox{Figure~\ref{ptau}}, via the gray bars on the right side.  As previously discussed by \ct{Seager05} and \ct{Fortney06b}, since the efficiency of energy redistribution and the depth to which one ``sees'' are wavelength dependent, the shape and times of maxima and minima in thermal emission light curves will be a function of wavelength.

Examination of \mbox{Figure~\ref{ptau}} shows that for a pM Class planet winds speeds of \emph{dozens} of km s$^{-1}$ would be necessary for advection to dominate over cooling/heating!  For a pL Class planet more modest winds speeds of several km s$^{-1}$ are needed.  Several km s$^{-1}$ winds are in line with predictions of a number of 2D and 3D dynamical models for hot Jupiter atmospheres \cp{Showman02,CS05,Langton07,Dobbs07}.  These calculated wind speeds are far below those needed to advect air before it radiatively heats/cools in a pM Class atmosphere. 

Winds will not be able to redistribute energy at the photospheres of pM Class planets.  The atmospheres of pM Class planets likely appear as one would expect from radiative equilibrium considerations:  the hottest part of the atmosphere is at the substellar point, and the atmosphere is cooler as one moves toward the planet's limb.  The night-side temperature will be relatively cold and will be set by the intrinsic flux from the interior of the planet, as well as the efficiency of energy redistribution at depth.  For a pL Class atmosphere, there will be a much more complex interplay between radiation and dynamics; energy redistribution may lead to a planetary hot spot being blown downwind.  The location of the hot spot will itself be wavelength dependent, and there will be increased energy transfer to the night side at photospheric pressures.  Variability in secondary eclipse depth is a possibility \cp{Rauscher07}.  Recently \ct{Dobbs07} have shown that larger day-night temperature differences are expected with increased atmospheric opacity, but they models lack motivation for their opacity choices.  Dynamical models that do not include a realistic treatment of opacities and radiative transfer will miss the important differences between these two classes of planets.

\section{Interiors}
We now have a collection of at least 20 transiting gas giant planets, a smaller planet that is perhaps Neptune-like (GJ 436b), and a potential hybrid planet (\hh, which is $\sim$2/3 heavy elements, by mass).  Very soon planets as small as 2 \re\ will be detected by COROT, and even smaller planets will be detected by Kepler.  It is worthwhile to examine expectations for all of these classes of planets, and how these expectations are being challenged by observations.

\subsection{Radii of Water-Rock-Iron Planets}
It seems likely that planets with masses within an order of magnitude of the Earth's mass will be composed primarily of more refractory species, like the planetary ices, rocks, and iron.  Within our solar system, objects of similar radius can differ by over a factor of 3 in mass, due to compositional differences.  A planet with the radius of Mercury, which is potentially detectable with Kepler, could indicate a mass of 0.055 \me, like Mercury itself, or a mass of 1/3 this value, like Callisto, which has a radius that differs by only 30 km.  \ct{Fortney07a} examined the radii of these planets, focusing on water/rock and rock/iron mixtures.  These include pure water and water/rock mixtures, which could be described as ``water worlds'' or ``Ocean planets."  \citet{Kuchner03} and \citet{Leger04} have pointed out water-rich objects could reach many Earth masses (perhaps as failed giant planet cores) and migrate inward to smaller orbital distances.  \ct{Fortney07a} also consider planets composed of pure rock, rock and iron mixtures, and pure iron, more similar to our own terrestrial planets.

\begin{figure}[!ht]
\begin{center}
\includegraphics[scale=0.75]{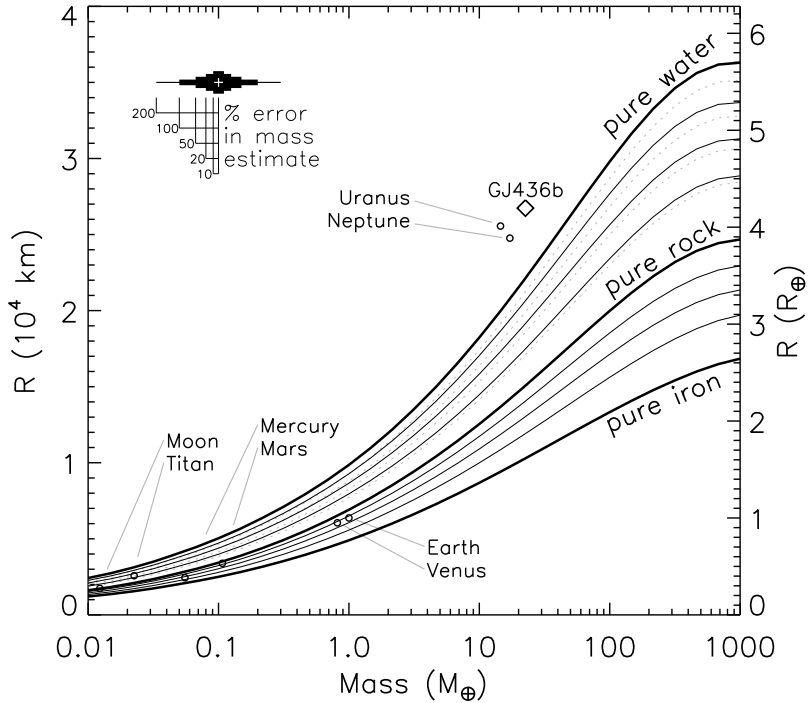}
\end{center}
\caption{Mass (in \me) vs.~radius (in km and \re) for planets composed for water, rock, and iron.  The topmost thick black curve is for pure ``warm'' water.  The middle thick curve is for pure rock (Mg$_2$SiO$_4$).  The bottommost thick curve is for pure iron (Fe).  The three black thin curves between pure water and pure rock, are from top to bottom, 75\% water/25\% rock, 50/50, and 25/75.  The inner layer is rock and the outer layer is water.  The gray dotted lines are the same pure water and water/rock curves, but for zero-temperature water.  The three black thin curves between pure rock and iron, are from top to bottom, 75\% rock/25\% iron, 50/50, and 25/75.  The inner layer is iron and the outer layer is rock.  At the upper left we show the horizontal extent of mass error bars, for any given mass.}\label{fig:rmiri}
\end{figure}

The results are shown in \mbox{Figure~\ref{fig:rmiri}}.  For all compositions, the radii initially grow as $M^{1/3}$, but at larger masses, compression effects become important.  As a greater fraction of the electrons become pressure ionized, the materials begin to behave more like a Fermi gas, and there is a flattening of the mass-radius curves near 1000 \me.  Eventually the radii shrink as mass increases, with radii falling with $M^{-1/3}$ \citep[see][]{Zapolsky69}.
At the top left of \mbox{Figure~\ref{fig:rmiri}} we also show the size of various levels of uncertainty in planetary mass, as a percentages of a given mass, from 10 to 200\%.  For instance, if one could determine the mass of a 1 \me~planet to within 50\%,  even a radius determination accurate to within 0.25 \re~would lead to considerable ambiguity concerning composition, ranging from 50/50 water/rock to pure iron.  The shallow slope of the mass-radius curves below a few \me~makes accurate mass determinations especially important for understanding composition.  We note that all models for water/rock/iron planets to date, be them simple \cp{Fortney07a,Seager07} or complex \citep{Ehrenreich06,Valencia06}, give quite similar mass-radius relations, such that detailed models (that account for temperature effects and phase changes) do not appear to be necessary for determining likely bulk planetary composition \cp{Seager07,Fortney07a}.

From \mbox{Figure~\ref{fig:rmiri}} it is clear that planet GJ 436b cannot be composed purely of water.  Ammonia and methane are slightly less dense than water, and may also be found within transiting planets, although they condense at colder temperatures.  These three species are often termed ``planetary ices'' even though they may be found in any phase, not necessarily solid.  It appears that GJ 436b must have a H/He envelope, like Uranus and Neptune.  Furthermore, its radius is consistent with it \emph{not} being the remnant of an evaporated gas giant \cp{Fortney07a}.  It is not possible to determine its composition in detail, as any water/rock ratio, with a variety of masses, with the remaining mass due to H/He, could lead to the observed radius.  It is thought that Uranus and Neptune are composed mainly of the planetary ices in the dense warm fluid phase.

Since radial velocity or astrometric followup for even smaller planets will be extremely difficult and time consuming, radii may have to suffice as a proxy for mass for some time.  Given an assumed composition, such as ``Earth-like," one could assign masses to terrestrial-sized transiting planets.  The distribution of planetary masses vs.~orbital distance and stellar type could then be compared with planet formation models.

\begin{figure}[!ht]
\begin{center} 
\includegraphics[scale=0.7]{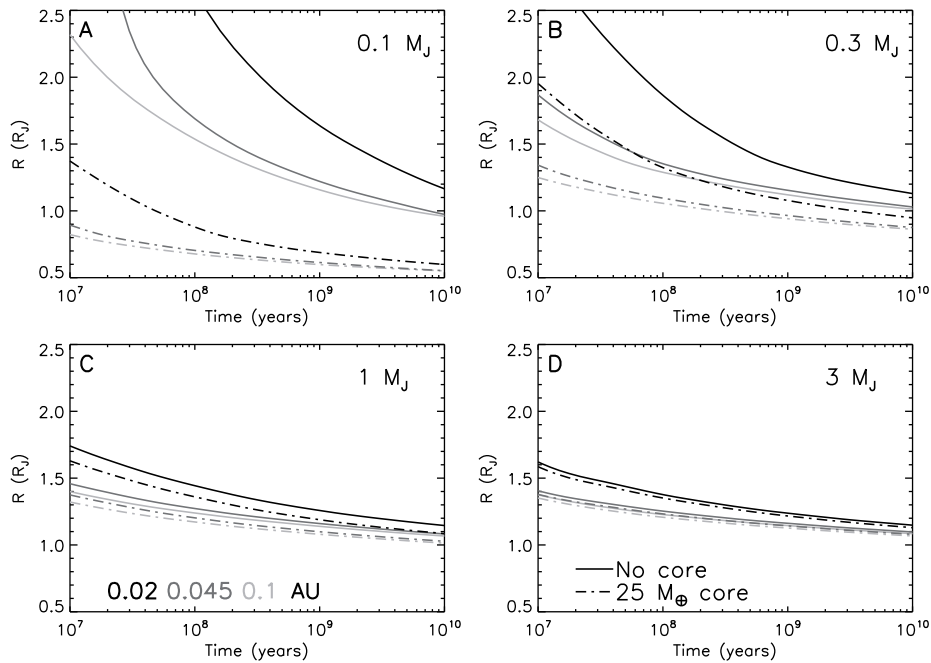}
\end{center}
\caption{Planetary radii as a function of time for masses of 0.1 \mj~(32 \me, $A$), 0.3 \mj~($B$), 1.0 \mj~($C$), and 3.0 \mj~($D$).  The three shades code for the three different orbital separations from the Sun, shown in ($C$).  Solid lines indicate models without cores and dash-dot lines indicate models with a core of 25 \me.}\label{fig:quad}
\end{figure}
\subsection{Radii of Gas Giants}
Planets around the mass of Uranus and Neptune ($\sim15$ \me) to objects as large at 75 \mj~can be described by the same cooling theory.  In general, planets with larger cores will have smaller radii, and planets closer to their parent stars will have larger radii at a given age then planets at larger orbital distances.  From transit observations we seek to understand the abundance of heavy elements within a giant planet's interior by comparing a measured mass/radius to theoretical calculations.  It is not possible to determine if these heavy elements are within a core or distributed in the H/He envelope.  Both Jupiter and Saturn have cores \emph{and} supersolar abundances of heavy elements in their H/He envelopes.  Even if Jupiter's central core is very small, from its mass/radius it must possess 20-30 \me\ of heavy elements somewhere within its interior.  Similar numbers are true for Saturn \cp{Saumon04}.  Given this knowledge, 1) it is very unlikely that the all heavy elements within a transiting giant planet are in the core, and 2) transiting planets (or any giant planets formed through core accretion) should not be expected to be pure H/He.

In \mbox{Figure~\ref{fig:quad}} we plot the contraction of planets from 0.1 to 3 \mj, as a function of age at various orbital distances \cp{Fortney07a}.  We also show the effect of a core of 25 \me, the approximate mass of heavy elements within Jupiter and Saturn \citep{Saumon04}.  Independent of mass, the spread in radii between 0.045 and 0.1 AU is small compared to the difference between 0.045 and 0.02 AU \cp[see][]{Fortney07a}.  As expected, the effect of the 25 \me~core diminishes with increased planet mass, as the core becomes a relatively smaller fraction of the planet's mass.  The radii at early ages it quite large, especially for the low-mass planets under intense stellar irradiation.  These hot low-gravity planets are potentially susceptible to evaporation \citep{Baraffe04, Hubbard07}.
\begin{figure}[!ht]
\begin{center}
\includegraphics[scale=0.67]{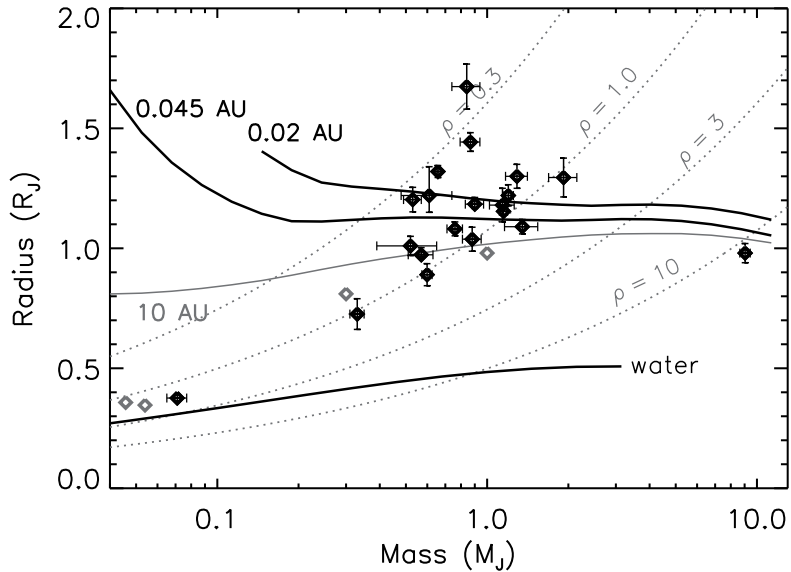}
\end{center}
\caption{Planetary radii from the models of \ct{Fortney07a} at 4.5 Gyr.  The models are for a composition of pure H/He (which is likely unrealistic given the structure of Jupiter \& Saturn), and includes the transit radius effect.  Gas giant models were calculated at 0.02, 0.045 and 10 AU from the Sun.  Diamonds without error bars or solar system planets.}\label{mr}
\end{figure}

\subsection{Current Explanations for Large Radii of Gas Giants}
The standard cooling theory for giant planets \cp[e.g.,][]{Hubbard02} envisions an adiabatic H/He envelope, likely enhanced in heavy elements, on top of a distinct heavy element core, likely composed of ices and rocks.  It is the radiative atmosphere that serves as the bottleneck for interior cooling and contraction.  As seen in \mbox{Figure~\ref{fig:quad}}, at 4.5 Gyr it is not possible to have radii much in excess of $\sim$1.2 \rj, even at 0.02 AU from the Sun, although at younger ages radii are larger.  Nevertheless, as seen in \mbox{Figure~\ref{mr}}, several planets have radii in excess of 1.2 \rj, and most receive irradiation far below that expected at 0.02 AU, as seen in \mbox{Figure~\ref{flux}}.  Explaining the large radii has been a major focus of exoplanet research for several years.  Below we briefly review the previous work.
\begin{itemize}
\item{\ct{Bodenheimer01} proposed that the radius of \hd\ could be explained by non-zero orbital eccentricity, forced by an unseen additional planetary companion.  This eccentricity would then be tidally damped, perhaps for gigayears.  This is potentially ruled out by the timing of the secondary eclipse \cp{Deming05b}.  But see also \ct{Mardling07}.}
\item{\ct{Guillot02} proposed that a small fraction ($\sim$1\%) of absorbed stellar flux is converted to kinetic energy (winds) and dissipated at a depth of tens of bars by, e.~g., the breaking of atmospheric waves.  This mechanism would presumably effect all hot Jupiters to some degree.}
\item{\ct{Baraffe04} found that \hd\ could be in the midst of extreme envelope evaporation, leading to a large radius, and we are catching the planet at a special time in its evolution.  The authors themselves judged this to be very unlikely.}
\item{\ct{Winn05} found that \hd\ may be stuck in a Cassini state, with its obliquity turned over at 90 degrees, which leads to a tidal damping of obliquity over gigayear ages.  Recently \ct{Levrard07} and \ct{Fabrycky07} have cast serious doubt on this mechanism.}
\item{\ct{Burrows07} propose that atmospheres with significantly enhanced opacities (10$\times$ that of a solar mixture) would stall the cooling and contraction of the planetary interior, leading to larger radii at gigayear ages.  Spectra of hot Jupiter atmospheres will either support or refute this (currently ad-hoc) possibility.}
\item{\ct{Chabrier07c}, independently following along the lines of a hypothesis from \ct{Stevenson85}, suggest that gradients in heavy elements (such as from core dredge-up or dissolution of planetesimals) could suppress convection and cooling in the H/He envelope, leading to large radii at gigayear ages.  The effect could be present at any orbital distance.}
\item{\ct{Hansen07} suggest that if mass loss due to evaporation leads to a preferential loss of He vs.~H (perhaps due to magnetic fields confining H$^+$), that planets could be larger than expected due to a smaller mean molecular weight.  This mechanism would also presumably effect all hot Jupiters to some degree. }
\end{itemize}

We note that a planetary embiggening mechanism that would affect all hot Jupiters is perfectly cromulent.  Since giant planets are expected to be metal rich (Jupiter and Saturn are 5-20\% heavy elements) a mechanism that would otherwise lead to large radii could easily be canceled out by a large planetary core or a supersolar abundance of heavy elements in the H/He envelope in most planets \cp{Fortney06}.  Planets that appear ``too small'' are certainly expected and are relatively easy to account for due to a diversity in internal heavy elements abundances \cp{Guillot06,Burrows07}.  It is important to recall that all these evolution models neglect (to the best of our knowledge) the strong opacity of TiO/VO on the hottest planets, which may alter their cooling history and and will lead to a larger optical transit radius \cp{Fortney07c,Burrows07c}.  This avenue will need to be explored.

\section{Conclusions}
The future will surely see additional discoveries in atmospheric and interior physics.  In particular, a large amount of \emph{Spitzer} data on atmospheric thermal emission is just beginning to be published.  Thermal emission light curves, as a function of orbital phase, will take on increased importance.  Mass/radius determinations for less irradiated giant planets will test theories for planetary evolution.  Certainly the next frontier is smaller masses, now that GJ 436b in known to transit \cp{Gillon07}.  Paced by discoveries from space missions (such as \emph{COROT} and \emph{Kepler}) and ground-based surveys, as well as precise followup characterization observations via \emph{Spitzer}, \emph{Hubble}, \emph{MOST}, and the ground, the future is even more exciting than the present.

\acknowledgements 
J.~J.~Fortney acknowledges the support of a Spitzer Fellowship at the NASA Ames Research Center and the SETI Institute.



\begin{thebibliography}{}

\bibitem[{{Baraffe} {et~al.}(2004){Baraffe}, {Selsis}, {Chabrier}, {Barman},
  {Allard}, {Hauschildt}, \& {Lammer}}]{Baraffe04}
{Baraffe}, I., {Selsis}, F., {Chabrier}, G., {Barman}, T.~S., et.~al 2004, \aap, 419, L13

\bibitem[{{Bodenheimer} {et~al.}(2001){Bodenheimer}, {Lin}, \&
  {Mardling}}]{Bodenheimer01}
{Bodenheimer}, P., {Lin}, D.~N.~C., \& {Mardling}, R.~A. 2001, \apj, 548, 466

\bibitem[{{Burrows} {et~al.}(2007{\natexlab{a}}){Burrows}, {Budaj}, \&
  {Hubeny}}]{Burrows08}
{Burrows}, A., {Budaj}, J., \& {Hubeny}, I. 2007{\natexlab{a}}, ApJ, submitted

\bibitem[{{Burrows} {et~al.}(2007{\natexlab{b}}){Burrows}, {Hubeny}, {Budaj},
  \& {Hubbard}}]{Burrows07}
{Burrows}, A., {Hubeny}, I., {Budaj}, J., \& {Hubbard}, W.~B.
  2007{\natexlab{b}}, \apj, 661, 502

\bibitem[{{Burrows} {et~al.}(2007{\natexlab{c}}){Burrows}, {Hubeny}, {Budaj},
  {Knutson}, \& {Charbonneau}}]{Burrows07c}
{Burrows}, A., {Hubeny}, I., {Budaj}, J., {Knutson}, H., \& {Charbonneau}, D.
  2007{\natexlab{c}}, ApJL in press, ArXiv e-prints/0709.3980

\bibitem[{{Burrows} {et~al.}(2000){Burrows}, {Marley}, \& {Sharp}}]{BMS}
{Burrows}, A., {Marley}, M.~S., \& {Sharp}, C.~M. 2000, \apj, 531, 438

\bibitem[{{Chabrier} \& {Baraffe}(2007)}]{Chabrier07c}
{Chabrier}, G. \& {Baraffe}, I. 2007, \apjl, 661, L81

\bibitem[{{Cooper} \& {Showman}(2005)}]{CS05}
{Cooper}, C.~S. \& {Showman}, A.~P. 2005, \apjl, 629, L45

\bibitem[{{Deming} {et~al.}(2005){Deming}, {Seager}, {Richardson}, \&
  {Harrington}}]{Deming05b}
{Deming}, D., {Seager}, S., {Richardson}, L.~J., \& {Harrington}, J. 2005,
  Nature, 434, 740

\bibitem[{{Dobbs-Dixon} \& {Lin}(2007)}]{Dobbs07}
{Dobbs-Dixon}, I. \& {Lin}, D.~N.~C. 2007, ApJ in press, astro-ph/0704.3269

\bibitem[{{Ehrenreich} {et~al.}(2006){Ehrenreich}, {Lecavelier des Etangs},
  {Beaulieu}, \& {Grasset}}]{Ehrenreich06}
{Ehrenreich}, D., {Lecavelier des Etangs}, A., {Beaulieu}, J.-P., \& {Grasset},
  O. 2006, \apj, 651, 535

\bibitem[{{Fabrycky} {et~al.}(2007){Fabrycky}, {Johnson}, \&
  {Goodman}}]{Fabrycky07}
{Fabrycky}, D.~C., {Johnson}, E.~T., \& {Goodman}, J. 2007, \apj, 665, 754

\bibitem[{{Fortney} {et~al.}(2006{\natexlab{a}}){Fortney}, {Cooper}, {Showman},
  {Marley}, \& {Freedman}}]{Fortney06b}
{Fortney}, J.~J., {Cooper}, C.~S., {Showman}, A.~P., {Marley}, M.~S., \&
  {Freedman}, R.~S. 2006{\natexlab{a}}, \apj, 652, 746

\bibitem[{{Fortney} {et~al.}(2007{\natexlab{a}}){Fortney}, {Lodders}, {Marley},
  \& {Freedman}}]{Fortney07c}
{Fortney}, J.~J., {Lodders}, K., {Marley}, M.~S., \& {Freedman}, R.~S.
  2007{\natexlab{a}}, ApJ, submitted

\bibitem[{{Fortney} {et~al.}(2007{\natexlab{b}}){Fortney}, {Marley}, \&
  {Barnes}}]{Fortney07a}
{Fortney}, J.~J., {Marley}, M.~S., \& {Barnes}, J.~W. 2007{\natexlab{b}}, \apj,
  659, 1661

\bibitem[{{Fortney} {et~al.}(2005){Fortney}, {Marley}, {Lodders}, {Saumon}, \&
  {Freedman}}]{Fortney05}
{Fortney}, J.~J., {Marley}, M.~S., {Lodders}, K., {Saumon}, D., \& {Freedman},
  R. 2005, \apjl, 627, L69

\bibitem[{{Fortney} {et~al.}(2006{\natexlab{b}}){Fortney}, {Saumon}, {Marley},
  {Lodders}, \& {Freedman}}]{Fortney06}
{Fortney}, J.~J., {Saumon}, D., {Marley}, M.~S., {Lodders}, K., \& {Freedman},
  R.~S. 2006{\natexlab{b}}, \apj, 642, 495

\bibitem[{{Gillon} {et~al.}(2007){Guillot}, {Santos}, {Pont}, {Iro}, {Melo},
  \& {Ribas}}]{Gillon07}
{Gillon}, M., {Pont}, F., et.~al 2007, \aap, 472, L13

\bibitem[{{Guillot} {et~al.}(2006){Guillot}, {Santos}, {Pont}, {Iro}, {Melo},
  \& {Ribas}}]{Guillot06}
{Guillot}, T., {Santos}, N.~C., {Pont}, F., {Iro}, N., {Melo}, C., \& {Ribas},
  I. 2006, \aap, 453, L21

\bibitem[{{Guillot} \& {Showman}(2002)}]{Guillot02}
{Guillot}, T. \& {Showman}, A.~P. 2002, \aap, 385, 156

\bibitem[{{Hansen} \& {Barman}(2007)}]{Hansen07}
{Hansen}, B.~M.~S. \& {Barman}, T. 2007, ApJ in press, ArXiv e-prints/0706.3052

\bibitem[{{Harrington} {et~al.}(2007){Harrington}, {Luszcz}, {Seager},
  {Deming}, \& {Richardson}}]{Harrington07}
{Harrington}, J., {Luszcz}, S., {Seager}, S., {Deming}, D., \& {Richardson},
  L.~J. 2007, Nature, 447, 691

\bibitem[{{Hubbard} {et~al.}(2002){Hubbard}, {Burrows}, \&
  {Lunine}}]{Hubbard02}
{Hubbard}, W.~B., {Burrows}, A., \& {Lunine}, J.~I. 2002, \araa, 40, 103

\bibitem[{{Hubbard} {et~al.}(2007){Hubbard}, {Hattori}, {Burrows}, {Hubeny}, \&
  {Sudarsky}}]{Hubbard07}
{Hubbard}, W.~B., {Hattori}, M.~F., {Burrows}, A., {Hubeny}, I., \& {Sudarsky},
  D. 2007, Icarus, 187, 358

\bibitem[{{Hubeny} {et~al.}(2003){Hubeny}, {Burrows}, \& {Sudarsky}}]{Hubeny03}
{Hubeny}, I., {Burrows}, A., \& {Sudarsky}, D. 2003, \apj, 594, 1011

\bibitem[{{Knutson} {et~al.}(2007){Knutson}, {Charbonneau}, {Allen}, {Burrows},
  \& {Megeath}}]{Knutson08}
{Knutson}, H.~A., {Charbonneau}, D., {Allen}, L.~E., {Burrows}, A., \&
  {Megeath}, S.~T. 2007, ApJ in press, ArXiv e-prints/0709.3984

\bibitem[{{Kuchner}(2003)}]{Kuchner03}
{Kuchner}, M.~J. 2003, \apjl, 596, L105

\bibitem[{{Langton} \& {Laughlin}(2007)}]{Langton07}
{Langton}, J. \& {Laughlin}, G. 2007, \apjl, 657, L113

\bibitem[{{L{\'e}ger} {et~al.}(2004){L{\'e}ger}, {Selsis}, {Sotin}, {Guillot},
  {Despois}, {Mawet}, {Ollivier}, {Lab{\`e}que}, {Valette}, {Brachet},
  {Chazelas}, \& {Lammer}}]{Leger04}
{L{\'e}ger}, A., {Selsis}, F., {Sotin}, C., {Guillot}, T., et.~al 2004, Icarus, 169, 499

\bibitem[{{Levrard} {et~al.}(2007){Levrard}, {Correia}, {Chabrier}, {Baraffe},
  {Selsis}, \& {Laskar}}]{Levrard07}
{Levrard}, B., {Correia}, A.~C.~M., {Chabrier}, G., {Baraffe}, et.~al 2007, \aap, 462, L5

\bibitem[{{Mardling}(2007)}]{Mardling07}
{Mardling}, R.~A. 2007, MNRAS in press, ArXiv/0706.0224

\bibitem[{{Rauscher} {et~al.}(2007){Rauscher}, {Menou}, {Cho}, {Seager}, \&
  {Hansen}}]{Rauscher07}
{Rauscher}, E., {Menou}, K., {Cho}, J.~Y.-K., {Seager}, S., \& {Hansen},
  B.~M.~S. 2007, \apjl, 662, L115

\bibitem[{{Saumon} \& {Guillot}(2004)}]{Saumon04}
{Saumon}, D. \& {Guillot}, T. 2004, \apj, 609, 1170

\bibitem[{{Seager} {et~al.}(2007){Seager}, {Kuchner}, {Hier-Majumder}, \&
  {Militzer}}]{Seager07}
{Seager}, S., {Kuchner}, M., {Hier-Majumder}, C., \& {Militzer}, B. 2007, ApJ
  in press, ArXiv e-prints/0707.2895

\bibitem[{{Seager} {et~al.}(2005){Seager}, {Richardson}, {Hansen}, {Menou},
  {Cho}, \& {Deming}}]{Seager05}
{Seager}, S., {Richardson}, L.~J., {Hansen}, B.~M.~S., et.~al 2005, \apj, 632, 1122

\bibitem[{{Showman} \& {Guillot}(2002)}]{Showman02}
{Showman}, A.~P. \& {Guillot}, T. 2002, \aap, 385, 166

\bibitem[{{Stevenson}(1985)}]{Stevenson85}
{Stevenson}, D.~J. 1985, Icarus, 62, 4

\bibitem[{{Valencia} {et~al.}(2006){Valencia}, {O'Connell}, \&
  {Sasselov}}]{Valencia06}
{Valencia}, D., {O'Connell}, R.~J., \& {Sasselov}, D. 2006, Icarus, 181, 545

\bibitem[{{Winn} \& {Holman}(2005)}]{Winn05}
{Winn}, J.~N. \& {Holman}, M.~J. 2005, \apjl, 628, L159

\bibitem[{{Yelle}(2004)}]{Yelle04}
{Yelle}, R.~V. 2004, Icarus, 170, 167

\bibitem[{{Zapolsky} \& {Salpeter}(1969)}]{Zapolsky69}
{Zapolsky}, H.~S. \& {Salpeter}, E.~E. 1969, \apj, 158, 809

\end{thebibliography}


\end{document}